\documentclass{iaus}
\usepackage{graphicx}

\title[JD 11.~~ Ultra-soft sources as  type Ia progenitors] 
{Ultra-soft sources as supernova type Ia progenitors}

\author[Kelly Lepo \& Marten van Kerkwijk]   
{Kelly Lepo
 \and Marten van Kerkwijk}

\affiliation{Department of Astronomy and Astrophysics, University of Toronto,\\ 50 St. George Street, Toronto, ON M5S 3H4, Canada \\ email: {\tt lepo@astro.utoronto.ca,
mhvk@astro.utoronto.ca}}

\pubyear{2011}
\volume{281}  
\pagerange{}
\jname{Binary Paths to Type Ia Supernovae Explosions}
\editors{R. Di Stefano \& M. Orio, eds.}
\begin{document}

\maketitle

\begin{abstract}
Missing from the usual considerations of nuclear burning white dwarfs as Type Ia supernova progenitors are systems with very higher mass transfer rates, where more material than is needed for steady burning accretes on the white dwarf. This will move out the photosphere of the white dwarf, causing it to emit at longer wavelengths. Thus, we propose the name ultra-soft source (USS) for these objects.

We present a VLT/FLAMES survey looking for USS in the SMC, selected to be bright in the far UV and with blue far UV-V colors.  While we find some unusual objects, and recover known planetary nebula and WR stars, we detect no objects with strong He II lines, which should be a signature of USS. This null result either puts an upper limit on the number of USS in the SMC, or shows that we do not understand what the optical spectra of such objects will look like.

We also discuss the unusual LMC [WN] planetary nebula LMC N66 as a possible example of an USS. It has a luminosity consistent with that expected, and its spectra show incompletely CNO-processed material --- strong helium lines, some hydrogen, enhanced nitrogen and depleted carbon. It also shows periodic outbursts. USS may resemble N66 in quiescence. However, it lacks a FUV excess, contrary to our predictions.

\end{abstract}

\firstsection 
\section{Introduction}

As outlined in \cite{nomoto07}, nuclear burning white dwarfs can be divided into three classes, depending on the accretion rate from the secondary onto the white dwarf ($\dot{M}$). Novas, white dwarfs (WDs) accreting below the steady nuclear burning rate ($\lesssim 1 \times 10^{-7} M_{\odot}{\rm\ yr^{-1}}$, depending on the mass of the white dwarf), will undergo periodic, explosive nuclear burning. 

Within a narrow band of accretion rates ($\sim 1-4 \times 10^{-7} M_{\odot} {\rm\ yr^{-1}}$, depending on the mass of the WD), stable nuclear burning occurs on the surface. Unlike in a nova, when fusion is ignited in its base, the thicker hydrogen shell can expand and lift degeneracy enough to allow stable burning. The effective temperature of the white dwarf places the peak of its black body spectrum in the soft X-rays, thus they are referred to as super soft sources (SSS, \cite[van den Heuvel, \etal\ 1992]{vdH1992}). Super soft sources are often considered one of the most promising single degenerate (SD) type Ia supernova progenitors. 

Less considered in discussions of nuclear burning white dwarfs are those that are accreting above the steady nuclear burning rate, but below the Eddington limit ($\sim 4 \times 10^{-7} -  1 \times 10^{-5} M_{\odot} {\rm\ yr^{-1}}$). These white dwarfs will have a hydrogen burning layer over the degenerate core, with an extended envelope. They will not be super-soft X-ray emitters, since the envelope is thick enough to absorb soft X-rays. The larger photosphere of such objects will move the peak emission into longer wavelengths, likely the (far) UV. For lack of a better name, we dub these ultra-soft sources (USS).

There are two general ideas for what a USS might look like. The first is that the white dwarf will have an extended envelope. Such an object may also lose mass from the L2 Lagrange point and/or form a common envelope (\cite[Nomoto \etal\ 1979]{nomoto79}, \cite[Iben 1988]{Iben88}). Most cataclysmic variables have periods of about 80 min to 10 hours, limiting the white dwarf's envelop to its $\sim1 R_{\odot}$ Roche lobe. If the system has not formed a common envelope, the white dwarf component will appear as a very small OB giant. 

The second possibility is that the white dwarf will lose mass through optically thick winds (\cite[Hachisu  \etal\  (1996)]{hachisu96}). Depending on the mass loss rate and the accretion rate, the white dwarf may or may not fill its Roche lobe. These may appear like Wolf-Rayet stars or [WR] planetary nebula. \cite{hachisu03a} and \cite{hachisu03b} use a similar model to explain V Sge and Vsge-like stars.  

\section{A hunt for USS in the SMC}
The Magellanic clouds are relatively close to us and have dense stellar fields. There is also a low column density between the Magellanic clouds and us, which will not absorb UV photons as happens in the Milky Way. They thus are ideal places to look for USS.

Because the Magellanic clouds are too bright for GALEX, we turn to a relatively obscure telescope, the Ultraviolet Imaging Telescope (UIT), which flew on two space shuttle missions in late 1990 and early 1995. UIT imaged both Magellanic clouds in several FUV filters, including B5 at 162 nm (\cite[Cornett \etal\ 1997]{cornett97}).  We redid the astrometry and cross-identified sources with the Magellanic Cloud Photometric Survey (MCPS, \cite[Zaritsky \etal\ 2002]{zaritsky04}) to get more accurate positions for sources in the UIT catalog. Since the UIT observations are cover the entire the central core of the SMC, while the LMC observations are in scattered fields, we chose to use the SMC data to simplify followup observations.

Using eqn. 4 from \cite{distefano10}, we can estimate the number of accreting WDs expected in the SMC, based on the SN Ia rate of the galaxy (as determined by its blue luminosity). Using similar values to Di Stafano, and  $L_B=4.4\times10^8$ for the SMC, we find that $N_{acc}$ (the number of accreting white dwarfs needed to reproduce the Type 1a rate) is 33--132 for the SMC. There are four known SSS in the SMC (\cite[Greiner 2000]{Grein00}). Assuming a single degenerate progenitor, there should be dozens missing progenitors in the SMC. While USS will probably not make up all of the missing progenitors (other good SD candidates include recurrent novae and symbiotic stars), white dwarfs with high accretion rates should exist. 

\begin{figure}[h]
\begin{center}
 \includegraphics[width= .69 \linewidth]{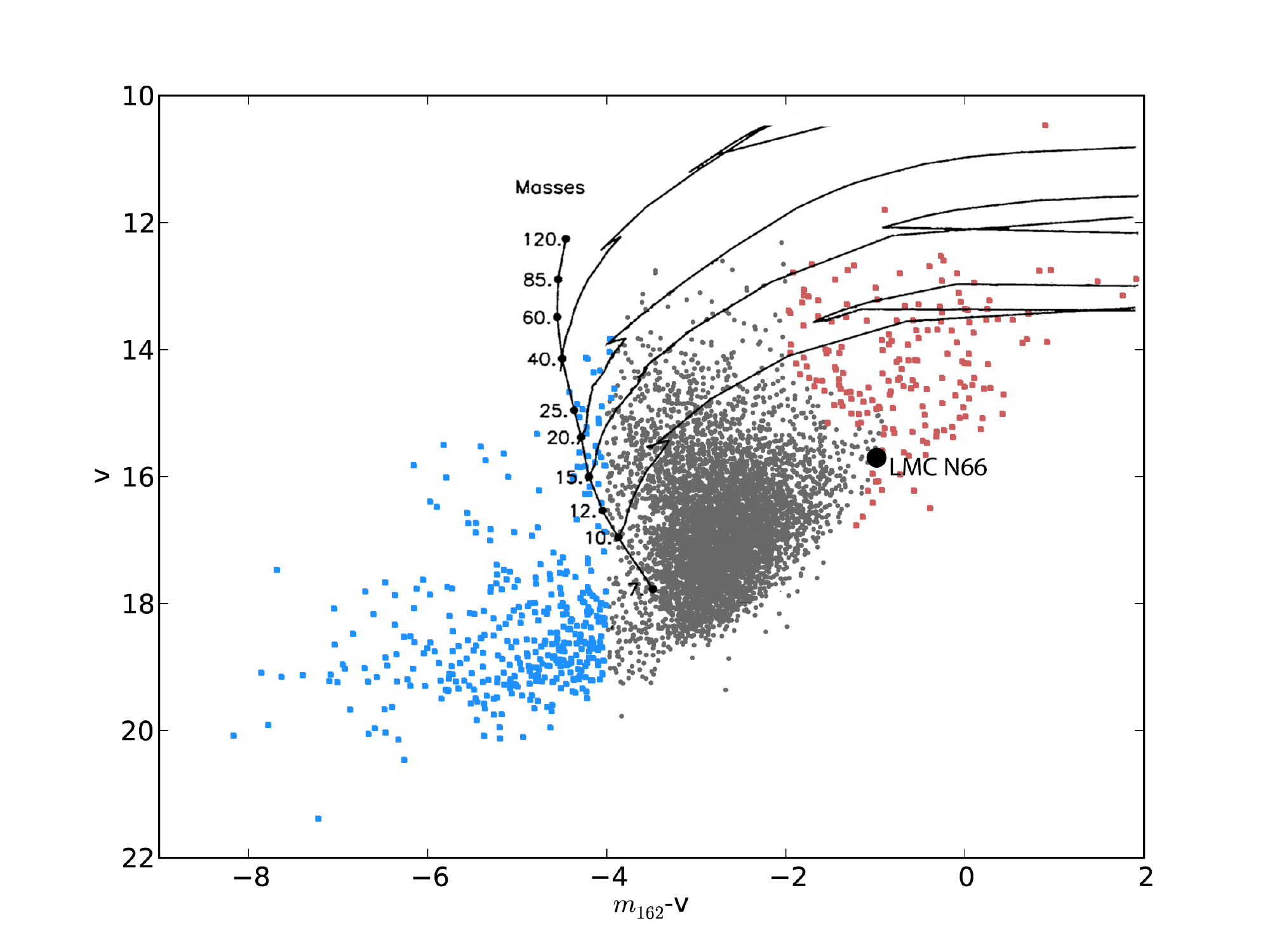}
 \caption{Color magnitude diagram of m$_{162}$-v vs v, using values from UIT and MCPS. There are 5723 points in the central clump (circles), 182 points redder than the central clump (squares) and 396 points bluer than the central clump (squares). Overlaid is fig. 4 from \cite[Cornett \etal\ (1997)]{cornett97}, showing evolutionary tracks of OB stars.  Also marked is the location of LMC N66, using values from the 25 April 1995 UIT observation and an April 1995 optical observation from \cite{pena96}. At least during outburst, LMC N66 does not have an UV excess.}
   \label{cmd}
\end{center}
\end{figure}

Figure \ref{cmd} is a color magnitude diagram of m$_{162}$-v vs v, using values from the FUV and optical catalogs. The blue edge of the central clump of stars corresponds to the position of the main sequence calculated by \cite[Cornett \etal\ (1997)]{cornett97}. Stars to the red side of the central clump are either evolved or reddened. Stars on the blue side of the central clump have an unexplained UV excess. This is the region where we expect to find USS. 

We took optical spectra of a majority of the objects outside of the central clump of fig. \ref{cmd}, using  the multi-object spectrometer VLT/FLAMES. We also took spectra of objects within the central clump using any extra fibers. We had five fields, which overlapped slightly, with $\sim 110$ objects in each field. Two fields were observed twice because the first observations were not taken within specified seeing conditions, but even the non-ideal observations turned out to be usable. In the end, we had 496 unique objects and 745 spectra. 

As expected, the majority of objects in the central Ômain sequenceÕ clump in fig. \ref{cmd}, were normal OB stars along with Be and B[e] stars. Stars to the red side of central clump (that had abnormally low FUV emission) were generally O, B and A giants.

Stars to the blue side of the central clump (with a FUV excess) were generally: (a) mismatches between the FUV and optical catalogs, (b) stars that seemed to have bad UVB measurements in the optical catalog, (c) objects that look like faint B stars, and (d) main sequence stars with nebular lines. Most of the main sequence stars were found near the edge of the central clump, so they likely belonged to the ``main sequence" group.

We did recover some known objects that are similar to what we expect a USS to look like: SMC-SMP 8 a [WC 8] planetary nebula with an unusual core, 2MASS J00520738-7235385 a (not particularly well studied) WR star, and several objects that look like planetary nebula. However, there were no objects with strong He II 4686 emission lines, which are found in the spectra of SSS, V Sge-like stars, and the LMC [WN] planetary nebula N66. It is likely that the hot WD in a USS would produce ionized helium emission lines, meaning it is unlikely we observed any USS. This null result either puts an upper limit on the number of USS in the SMC, or shows that we do not understand what the optical spectra of such objects will look like.

\section{LMC N66: A possible USS?}
LMC N66 (also known as WS 35 and SMP 83) is a unusual early [WN] planetary nebula located in the LMC. It undergoes periodic outbursts, where its luminosity increases by about an order of magnitude for a period of years. Its pre-outburst luminosity ($\log[L/L_{\odot}]= 4.5$) is higher than known [WC] planetary nebulae, and its outburst luminosity ($\log[L/L_{\odot}]= 5.4$) is on par with galactic WN stars (\cite[Hamann \etal\ 2003]{hamann03}).

Two outbursts have been observed,  with peak luminosities occurring in 1994 and 2007 (\cite[Pe\~{n}a \etal\ 1995]{pena95}, \cite[Hamann \etal\ 2003]{hamann03}, \cite[Pe\~{n}a \etal\ 2008]{pena08}). There no evidence in archival data that any outbursts occurred from 1955 to 1990 (\cite[Hamann \etal\ 2003]{hamann03}). 

Spectra of LMC N66 show incompletely CNO-processed material --- strong helium emission lines, some hydrogen, enhanced nitrogen and depleted carbon. From the width of the lines one infers with a low terminal wind speed. The model that has the fewest contradictions with the observed properties LMC N66, is that the central object of the planetary nebula is a low-mass binary, with a WD primary and a non-degenerate secondary. The WD rapidly accretes mass, and loses some to winds. The outbursts --- hard to explain with a single star model --- may be due to helium shell flashes on the white dwarf (\cite[Hamann \etal\ 2003]{hamann03}). 

Assuming the system is a white dwarf and a normal star, models by \cite[Hamann \etal\ (2003)]{hamann03} put the radius of the photosphere of the WD at $0.52-1.38 R_{\odot}$. This is about what one would expect for the Roche lobe of a white dwarf in a close binary. They find the mass loss rate of the object varies from $\dot{M} = 10^{-5.7}$ during quiescence to $\dot{M} =10^{-5.0} M_{\odot} {\rm\ yr^{-1}}$ during outburst. This is would put the white dwarf in the high $\dot{M}$ regime. Both the radius and $\dot{M}$ calculations are consistent with our ideas of USS.

LMC N66 was observed by UIT on 25 April 1995, but it was so close to the edge of the photographic plate that is was not included in the catalog by \cite{parker98}. We re-did the photometry of the entire UIT LMC field that LMC N66 was in, scaled our uncalibrated magnitudes so that they matched the published, calibrated ones, and then calculated the magnitude of LMC N66. We find m$_{162} = 15.22$. When UIT observed LMC N66, the nebula had passed the peak of its outburst, but was not yet in quiescence. \cite{pena96} also observed LMC N66 in April 1995 at optical wavelengths and find $m_{v}=16.2 \pm 0.2$. 

Putting those two observations together and plotting on fig. \ref{cmd}, we see that LMC N66 falls to the red side of the central clump, meaning than instead of having a UV excess, as was expected, the object has less FUV emission than main sequence stars. However, since we only have one observation during the end of an outburst, this may not be applicable during quiescence or other parts of the outburst cycle. In addition, since LMC N66 was near the edge of the plate, its flux may have been underestimated.

Although \cite[Hamann \etal\ (2003)]{hamann03} suggest that the best model of LMC N66 is for a a white dwarf binary with a low mass companion, very little work has been done to verify if the core of LMC N66 is in fact a binary. High resolution spectroscopy to look for radial velocity variations and photometry to look for $\sim 1$ day variations in the light curve might help to confirm the binary model.

\end{document}